# Deep learning-based reconstruction of highly accelerated 3D MRI

Sangtae Ahn, Uri Wollner, Graeme McKinnon, Isabelle Heukensfeldt Jansen, Rafi Brada, Dan Rettmann, Ty A. Cashen, John Huston, III, J. Kevin DeMarco, Robert Y. Shih, Joshua D. Trzasko, Christopher J. Hardy, and Thomas K. F. Foo

**Purpose**: To accelerate brain 3D MRI scans by using a deep learning method for reconstructing images from highly-undersampled multi-coil k-space data

**Methods**: DL-Speed, an unrolled optimization architecture with dense skip-layer connections, was trained on 3D T1-weighted brain scan data to reconstruct complex-valued images from highly-undersampled k-space data. The trained model was evaluated on 3D MPRAGE brain scan data retrospectively-undersampled with a 10-fold acceleration, compared to a conventional parallel imaging method with a 2-fold acceleration. Scores of SNR, artifacts, gray/white matter contrast, resolution/sharpness, deep gray-matter, cerebellar vermis, anterior commissure, and overall quality, on a 5-point Likert scale, were assessed by experienced radiologists. In addition, the trained model was tested on retrospectively-undersampled 3D T1-weighted LAVA (Liver Acquisition with Volume Acceleration) abdominal scan data, and prospectively-undersampled 3D MPRAGE and LAVA scans in three healthy volunteers and one, respectively.

**Results**: The qualitative scores for DL-Speed with a 10-fold acceleration were higher than or equal to those for the parallel imaging with 2-fold acceleration. DL-Speed outperformed a compressed sensing method in quantitative metrics on retrospectively-undersampled LAVA data. DL-Speed was demonstrated to perform reasonably well on prospectively-undersampled scan data, realizing a 2-5 times reduction in scan time.

**Conclusion**: DL-Speed was shown to accelerate 3D MPRAGE and LAVA with up to a net 10-fold acceleration, achieving 2-5 times faster scans compared to conventional parallel imaging and acceleration, while maintaining diagnostic image quality and real-time reconstruction. The brain scan data-trained DL-Speed also performed well when reconstructing abdominal LAVA scan data, demonstrating versatility of the network.

## I. Introduction

ACHIEVING faster MRI scans by accelerating or undersampling *k*-space data necessary to reconstruct an image has been a long-standing but important challenge. Fast MRI scans contribute to improved patient comfort and experience, in addition to allowing for higher patient throughput. MRI acceleration techniques also enable better spatial and temporal resolution and a larger FOV without increasing scan times. In addition, faster MRI scans are robust to patient motion, mitigating motion-induced artifacts that require repeat scans.[1]

Parallel imaging[2-4] and compressed sensing (CS)[5] are widely used to reconstruct MR images from undersampled k-space data for scan acceleration. Recently, deep learning (DL) techniques have been shown to outperform CS in reconstructing undersampled MR data and become a subject of great interest in the MRI community.[6] Handcrafted sparsity-enforcing regularization functions such as total variation used in CS are suboptimal for sparse MR reconstruction, often yielding blocky unnatural images particularly for low SNR and highly undersampled data. On the other hand, DL methods can learn data-driven regularization, leading to better image quality and lower reconstruction errors. Whereas it is a pain point to determine hyper-parameters in CS, DL model parameters can be learned from training data. In addition, DL inference can be efficiently deployed on a graphics processing unit (GPU), usually faster than CS reconstruction. Recently, Li et al. demonstrated a fast reconstruction method using generalized adversarial networks for whole-brain imaging.[7] However, they were only able to achieve up to a 6-fold acceleration.

In this paper, we present a DL-based unrolled optimization method, called DL-Speed, which is a densely connected iterative network,[8] for reconstructing complex-valued MR images from Cartesian-undersampled multi-coil k-space data and evaluate its performance on various datasets. Since the variational network, the first unrolled optimization technique applied to sparse MRI reconstruction, was proposed by Hammernik et al.,[9] many unrolled optimization variants have been investigated for MRI acceleration.[6, 10-22] Unrolled optimization is model-based or physics-based in the sense that a data-formation physics model that maps an image into

This work was supported in part by CDMRP under Grant W81XWH-16-2-0054 and by NIH under Grant U01EB024450.
S. Ahn is with GE Research, Niskayuna, NY, USA (e-mail: ahns@ge.com).
I. Heukensfeldt Jansen, C. J. Hardy and T. K. F. Foo are with GE Research, Niskayuna, NY, USA.
U. Wollner and R. Brada are with GE Research, Herzliya, Israel.
G. McKinnon is with GE Healthcare, Waukesha, WI, USA.
D. Rettmann is with GE Healthcare, Rochester, MN, USA.
T. A. Cashen is with GE Healthcare, Madison, WI, USA.
J. Huston and J. D. Trzasko are with Mayo Clinic, Rochester, MN, USA.
J. K. DeMarco is with Walter Reed National Military Medical Center, Bethesda, MD, USA.
R. Y. Shi is with Walter Reed National Military Medical Center, Bethesda, MD, USA, and Uniformed Services University, Bethesda, MD, USA.

undersampled k-space data, incorporating spatial coil sensitivity maps. The Fourier transform and undersampling is explicitly included in the reconstruction process. Other non-model-based DL approaches for accelerating scans include those that directly map undersampled k-space data into reconstructed images,[23] and image-based DL methods that map undersampling-artifact corrupted images into artifact-free images.[24] Model-based unrolled optimization methods do not need to learn the data formation process and therefore will require a lower complexity and a smaller training dataset in principle provided that the physics model is accurate.[10]

DL-Speed, an unrolled optimization method we investigate in this paper, has the following novel features. First, the architecture has skip-layer connections[25] between each iteration and a number of previous iterations (e.g., 10-20 skip connections). The dense skip connections strengthen feature propagation, making the network more robust, and have been shown to improve image quality in sparse MRI reconstruction.[8] Second, a contrast-weighted structural similarity index measure (SSIM)[26, 27] extended to complex-valued images is used as a loss function. A contrast-weighted SSIM loss function has been shown to yield sharper images than the commonly used L1 loss function.[26] All images remain complex-valued throughout the reconstruction process as well as when used by the loss function in training. Third, rather than using fully sampled data for training the neural networks,[10, 11, 13] we acquired mildly-undersampled data (e.g., an acceleration factor of about 2 or less) for training.[17] Mildly-undersampled data are easier to acquire (e.g., from a standard clinical protocol or a similar kind, and with breath-holding when needed), making it easier and faster to amass a large training dataset. In this study, we evaluate DL-Speed trained using mildly-undersampled data as a baseline, and artificially undersampling the data to arrive at higher acceleration factors.

The contribution of this paper is that we evaluate the entire pipeline of DL-Speed from data acquisition to image reconstruction for accelerated 3D brain MRI in terms of clinical acceptability and practicality. First, we focus on 3D MRI in this study, even though it is more straightforward to apply the methods to 2D. 3D MRI enables thinner slices and higher SNR at the expense of long scan times. Hence, a pressing need for faster, higher spatial resolution scans. Second, we evaluate image quality based on neuroradiologists' image interpretation as well as on objective metrics. Although useful, objective metrics of image quality, such as mean squared error (MSE) and SSIM[27], do not always correlate well with expert image assessment and human visual perception.[13] Therefore, our neuroradiologists' evaluation of diagnostic image quality is essential for assessing clinical acceptability.[13, 14, 17] Third, we evaluate DL-Speed on prospectively undersampled data as well as retrospectively undersampled data. A retrospective study is useful because reference images are available and therefore quantitative accuracy can be evaluated, on a one-to-one basis, with a known standard-of-reference. On the other hand, a prospective study is necessary for assessing the entire workflow including data acquisition, and it incorporates temporal acquisition order dynamics that impact image contrast. To our knowledge, there are few studies in the literature that test DL methods on prospectively undersampled data. Fourth, we pay attention to reconstruction times regarding the feasibility of real-time applications. Fifth, we evaluate the generalizability of DL-Speed by testing on 3D LAVA (Liver Acquisition with Volume Acceleration) abdominal scan data as well as 3D MPRAGE brain scan data. This allows us to assess the performance of the network in anatomies (e.g., the liver) that are quite distinct from that used in training the network (i.e., brain T1 and T2-weighted images). The combination of the above altogether makes a unique contribution to the field.

As the methods were refined and optimized, we have reported preliminary results in abstracts presented at International Society of Magnetic Resonance in Medicine Annual Meetings between 2018 and 2022.[8, 21, 26, 28]

## II. METHODS

All subject data used in this study were acquired under an IRB-approved protocol and written informed consent was provided. Images were acquired on 3.0 T whole-body (GE SIGNA Architect, GE Healthcare, Waukesha, WI, USA) and on a prototype 3.0 T head-only MRI system (MAGNUS, GE Research, Niskayuna, NY, USA)[29], using a 32-channel receive coil (NOVA Medical, Wilmington, MA, USA) unless stated otherwise.

### A. Network Architecture

Given Cartesian-undersampled multi-coil k-space data, $y$, a complex-valued image is reconstructed using an unrolled optimization scheme as a densely connected iterative network[8] that we now refer to as DL-Speed (see Figure 1 for a description of the architecture):

$$x_n = x_{n-1} - \lambda_n \mathcal{A}^*(y - \mathcal{A}x_{n-1}) - \mathcal{N}_{\theta_n}(x_{n-1}, x_{n-2}, \ldots, x_{\max(n-(G+1),1)})$$

for $n = 1, \ldots, N$ where $x_n$ is an intermediate complex-valued image after $n$ iterations, $\mathcal{A}$, is a linear forward operator that maps a complex-valued image into undersampled multi-coil k-space data, $\mathcal{A}^*$, is the adjoint of $\mathcal{A}$, $\lambda_n$ is a learnable parameter, $\mathcal{N}_\theta$, denotes a convolutional neural network with learnable network parameters $\theta$, and $N$ is the number of iterations. A zero-filled image was used as the initial image, $x_0$, and the $N$th iteration image, $x_N$, was taken as a final reconstructed image. The above equation is inspired by a gradient descent algorithm for regularized least squares[13] where the $\mathcal{A}^*(y - \mathcal{A}x)$ term is the gradient of the residual sum of squares $\|y - \mathcal{A}x\|^2$, hence called a data consistency term, $\lambda_n$, can be viewed as a step size of the gradient algorithm, and the neural network, $\mathcal{N}_\theta$, can be thought of as the gradient of a regularization function, which is learned from training data.

The forward operator, $\mathcal{A}$, in the data consistency term consists of 1) multiplying by coil sensitivity maps, 2) taking the Fourier transform, and 3) undersampling. The coil sensitivity maps were estimated using either a calibration scan or a fully-sampled central region of the $k$-space, and were normalized such that the sums of the squared magnitudes are all ones. For undersampling, a variable-density Poisson-disc (VDPD) sampling pattern[30] was used (see Supporting Information Figure 1S for an example). Pseudo-random variable-density sampling is widely used in DL[10, 11, 13, 14, 18] as well as CS[5] both for 2D and 3D.

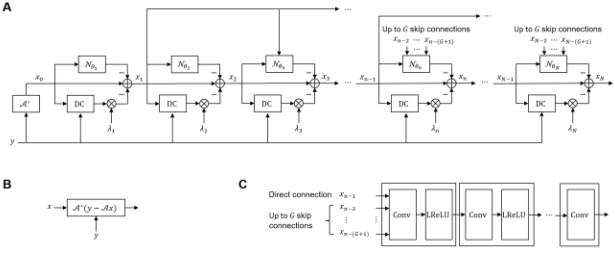

Figure 1. DL-Speed architecture. A: DL-Speed consists of multiple ($N$) iterations, each of which has a data consistency (DC) unit and a regularization unit based on a convolutional neural network $\mathcal{N}_{\theta_n}$. The DL-Speed generates a reconstructed image $x_N$ from undersampled multi-coil k-space data $y$. The multi-channel input to each neural network $\mathcal{N}_{\theta_n}$ includes not only the immediate past iterate $x_{n-1}$ but also up to $G$ further past iterates, $x_{n-2}, \ldots, x_{n-(G+1)}$, that is, up to $G$ skip-layer connections. In the diagram, $\lambda_n$ are learnable weight parameters, $\mathcal{A}$ denotes a forward model that maps an MR image into undersampled multi-coil k-space data, involving coil sensitivities, Fourier transform, and undersampling, and $\mathcal{A}^*$ is the adjoint of $\mathcal{A}$. B: Data consistency (DC) unit. C: Regularization unit, with a multi-channel input of up to $G+1$ images, $x_{n-1}, \ldots, x_{n-(G+1)}$, which consists of multiple convolutional (Conv) layers with leaky rectified liner unit (LReLu) activations.

The convolutional neural network consists of multiple convolutional layers with the leaky rectified linear unit (ReLU) activation (Figure 1C). The regularization unit, $\mathcal{N}_{\theta_n}$, that outputs the $n$th iteration image, $x_n$, takes as a multi-channel input not only the immediate previous iteration image, $x_{n-1}$ (i.e., a direct connection), but also up to $G$ further previous iterate images, $x_{n-2}, \ldots, x_{\max(n-(G+1),1)}$ (i.e., skip-layer connections) where $G$ is the number of skip connections. In the convolutional layers, the real part and the imaginary part of a complex-valued image were processed by separate channels.

### B. Loss Function

SSIM,[27] which is usually applied to real-valued nonnegative images, was extended to complex-valued images. The SSIM function for complex-valued images is defined as

$$\text{SSIM}_C(x,z) = [l_C(x,z)]^\alpha [c_C(x,z)]^\beta [s_C(x,z)]^\gamma$$

for two complex-valued image patches $x$ and $z$, where $l_C$, $c_C$ and $s_C$ are the luminance, contrast and structure comparison functions extended to complex-valued images, and $\alpha$, $\beta$ and $\gamma$ are weights. The comparison functions for complex-valued images are defined as

$$l_C(x,z) = \frac{1}{2}\left(\frac{2\text{Re}\{\mu_x\mu_z^*\} + c_1}{|\mu_x|^2 + |\mu_z|^2 + c_1} + 1\right)$$

$$c_C(x,z) = \frac{2\sigma_x\sigma_z + c_2}{\sigma_x^2 + \sigma_z^2 + c_2}$$

$$s_C(x,z) = \frac{|\sigma_{xz}| + c_3}{\sigma_x\sigma_z + c_3}$$

where $\text{Re}\{\cdot\}$ denotes the real part, $\mu_{(\cdot)}$ denotes the mean, that is, $\mu_x = (\sum_{i=1}^n x_i)/n$, $(\cdot)^*$ denotes the complex conjugate, $\sigma_{(\cdot)}$ denotes the standard deviation, that is, $\sigma_x = \sqrt{\mu_{|x-\mu_x|^2}}$, $\sigma_{(\cdot)(\cdot)}$ denotes the covariance, that is, $\sigma_{xz} = \mu_{(x-\mu_x)(z-\mu_z)^*}$, and $c_1$, $c_2$ and $c_3$ are positive constants such that $c_1 = (0.01L)^2$, $c_2 = (0.03L)^2$ and $c_3 = c_2/2$ as used by Wang et al.[27] with $L$ being the dynamic range. If the image patches are magnitude images, $c_C$ and $s_C$ becomes the same as the original contrast and structure comparison functions[27] for nonnegative images, respectively, and $l_C$ becomes affine in the original luminance comparison function[27], $l$, for nonnegative images, that is, $l_C = (l+1)/2$. Therefore, when $x$ and $z$ are magnitude images, $\text{SSIM}_C$ essentially ends up as the original SSIM[27] for nonnegative images.

Given a network output image and a reference target image, for each pixel, 11×11 image patches centered about the pixel from the two images are used to calculate $1 - \text{SSIM}_C$, and the mean of the $1 - \text{SSIM}_C$ values over the entire image pixels was used as the loss function.

Usually, unweighted SSIM with $\alpha = \beta = \gamma = 1$ is used. Weighting the contrast comparison function, i.e., $\beta > \alpha, \gamma$, tends to make reconstructed images sharper and less blurred.[26] A combination of the weights, $\alpha = 0.3$, $\beta = 1$, and $\gamma = 0.3$, was empirically chosen for yielding sharper and less blurred images without creating artifacts and excessive noise.[26] In this paper, the contrast-weighted version of SSIM extended to complex-valued images with the combination of the weights was used for the loss function in DL-Speed training.

### C. Training

For training, 148 datasets of T1-weighted 3D BRAVO (GE Healthcare) brain scans acquired with an acceleration of 2 or less using a regular sampling pattern were collected. From each of the mildly-undersampled datasets, fully-sampled k-space data were regenerated using a parallel imaging technique, Autocalibrating Reconstruction for Cartesian imaging (ARC)[31]. Retrospectively-undersampled k-space data from the fully-sampled data with a 10-fold acceleration and a VDPD sampling pattern, a fully-sampled image as a target, and coil sensitivity maps estimated using either a calibration scan or a fully-sampled central region of the k-space were used for training. Out of the 148 datasets, 133 were used for training, and 15 for validation. DL-Speed was trained using TensorFlow (Google AI, Mountain View, CA) with a batch size of 1, and Adam optimizer[32] with a learning rate of $10^{-4}$.

### D. DL-Speed Variants

Three DL-Speed variants were trained using different types of filters in the regularization unit. In the first variant, called DL-Speed 2D Conv, 2D filters (3×3) were applied in the 2D plane defined by the phase encoding ($y$) and slice encoding ($z$) directions in the convolutional layers. For DL-Speed 2D Conv, 28 iterations ($N$), 9 convolutional layers per iteration, 96 filters per layer within each regularization unit, and up to 20 skip connections ($G$) were used. In the second variant, called DL-Speed 3D Conv, 3D filters (3×3×3) were used in the convolutional layers. In the third variant, called DL-Speed with alternating 2D Conv, 2D filters (3×3) were applied to the $yz$ plane for 3 iterations and to the $xy$ plane for 1 iteration, alternatingly. For DL-Speed 3D Conv and with alternating 2D

Conv, 10 iterations ($N$), 9 convolutional layers per iteration, 32 filters per layer, and up to 10 skip connections ($G$) were used. The complexity of DL-Speed 3D Conv and with alternating 2D Conv had to be lower than that of DL-Speed 2D Conv due to the memory limitation.

*E. Evaluation on MPRAGE Brain Data*

Sagittal 3D T1-weighted MPRAGE brain data (1.0-mm isotropic resolution) were acquired on a high-performance 3.0 T MAGNUS head-gradient system (200 mT/m and 500 T/m/s gradients)[29]. 3D MPRAGE scans, undersampled with a net 2.1-fold acceleration and a regular sampling pattern following a standard protocol, were acquired in 5 healthy volunteers for a retrospective undersampling study. Fully sampled k-space data were generated from the acquired k-space data via the ARC reconstruction, and were then retrospectively undersampled with a 10-fold acceleration using a VDPD sampling pattern.

DL-Speed images were reconstructed from the 10-fold retrospectively-undersampled data using the DL-Speed variants with 2D Conv, alternating 2D Conv and 3D Conv and compared to a baseline image reconstructed from the 2.1-fold undersampled data by ARC using MSE and SSIM. For each subject, 4 reconstructed image volumes (ARC and DL-Speed variants with 2D Conv, alternating 2D Conv and 3D Conv) were randomly arranged and scored by 3 board-certified neuroradiologists who were blinded to reconstruction methods. Image scoring was based on consensus reading using a 5-point Likert scale in 8 categories: SNR, artifacts, gray/white matter contrast, resolution/sharpness, deep gray, cerebellar vermis, anterior commissure, and overall quality. For the Likert scale, 1 was considered as poor, and 5 was excellent. The assessments of the deep gray-matter, cerebellar vermis, and anterior commissure categories focused on scoring anatomical structures with fine features that were thought to be vulnerable to artifacts, such as blurring.

In addition, 3D MPRAGE data prospectively undersampled with a 10-fold acceleration and a VDPD sampling pattern were acquired in further 3 volunteers and used for evaluating DL-Speed. Scans were acquired with the standard ARC protocol (with 2.1-fold acceleration), and a prospectively-undersampled DL-Speed protocol (with 10-fold acceleration).

*F. Evaluation on LAVA Abdominal Data*

Fifteen 3D T1-weighted LAVA abdominal scan datasets acquired with an acceleration factor of ≤2 using a regular sampling pattern on a 3.0 T whole-body system were collected. For each dataset, fully-sampled k-space data were generated by ARC and were retrospectively undersampled with an acceleration factor of about 9.4 using a pseudo-random VDPD sampling pattern (mean acceleration factor: 9.4, standard deviation: 0.3). From the retrospectively undersampled k-space data, images were reconstructed by DL-Speed and a total-variation based compressed sensing (CS) method that employs data-drive iterative soft thresholding,[33] and then compared against baseline ARC images.

For assessing performance with different anatomies (i.e., liver), breath-hold axial 3D T1-weighted LAVA scan data for the upper abdomen of a subject without contrast, prospectively undersampled with a net 9.3-fold acceleration using a VDPD sampling pattern were acquired on a 3.0 T whole-body system (FOV: 42.0×37.8 cm, slice thickness: 4 mm, frequency/phase encodings: 300×180, acquired slices: 48, scan time: 4.9 s, 18-channel data) with intermittent adiabatic fat saturation, in addition to a standard protocol scan with a net 3.8-fold acceleration and a regular sampling pattern (scan time: 11.8 s). The scanning procedure was then duplicated except with a pencil beam navigator tracker placed across the diaphragm to prospectively gate the acquisition during the exhalation phase of free breathing. Respiratory gating enables free-breathing scans, and the purpose here is to investigate the performance of DL-Speed on navigated scan data. Estimated navigated scan times for the standard regular-sampling scan with a 3.8-fold acceleration and the VDPD sampling scan with a 9.3-fold acceleration were 41.7 s and 16.4 s, respectively. The DL-Speed reconstructed images (9.3-fold acceleration) were subjectively assessed compared to the ARC reconstructed images (3.8-fold acceleration). We note that the assessment of a single subject for the liver data was to assess how well a brain-trained reconstruction network performed when used in a vastly different anatomical structure.

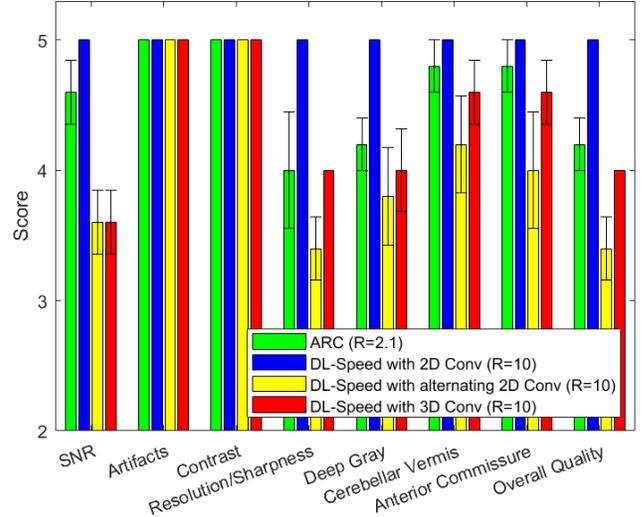

Figure 2. Radiologists' scores averaged over 5 subjects for ARC (net acceleration factor R=2.1) and DL-Speed with 2D Conv, alternating 2D Conv, and 3D Conv (net R=10), evaluated on retrospectively-undersampled 3D MPRAGE brain data, in 8 categories of SNR, artifacts, gray/white matter contrast, resolution/sharpness, deep gray, cerebellar vermis, anterior commissure, and overall quality. The error bars denote the standard errors.

III. RESULTS

*A. Evaluation on MPRAGE Brain Data*

Figure 2 shows the neuroradiologists' scores for the retrospectively-undersampled MPRAGE images, comparing the standard ARC (net acceleration factor R=2.1) and 3 DL-Speed variants (R=10) with 2D Conv, alternating 2D Conv, and 3D Conv. DL-Speed with 2D Conv received scores higher than or equal to those for the other methods including the baseline ARC in all 8 categories. All the methods were free of artifacts and showed excellent gray/white matter contrast. Figures 3 and 4 show example reconstructed images for comparison of the baseline ARC (R=2.1) and the 3 DL-Speed variants (R=10). DL-Speed with 2D Conv showed better recovery of the cerebellar vermis (indicated by the arrow in Figure 3) and the deep gray region (Figure 4) than the other methods. Table 1 shows the normalized MSE and SSIM for the DL-Speed variants compared to the standard ARC evaluated on the retrospectively-undersampled MPRAGE data (R=10). DL-Speed with 2D Conv outperformed DL-Speeds with alternating 2D Conv and 3D Conv in radiologists' scores (Figure 2) and normalized MSE (Table 1), and showed slightly lower yet comparable SSIM values (Table 1). Therefore, we chose DL-Speed with 2D Conv out of the DL-Speed variants and hereafter only showed results from DL-Speed with 2D Conv.

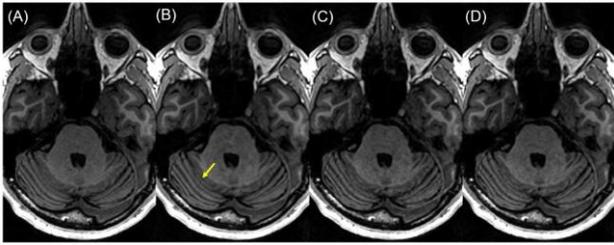

Figure 3. Example image slices, showing the cerebellar vermis indicated by the yellow arrow, reconstructed by (A) ARC (net acceleration factor R=2.1), (B) DL-Speed with 2D Conv (net R=10), (C) DL-Speed with alternating 2D Conv (net R=10), and (D) DL-Speed with 3D Conv (net R=10), evaluated on retrospectively-undersampled 3D MPRAGE brain data.

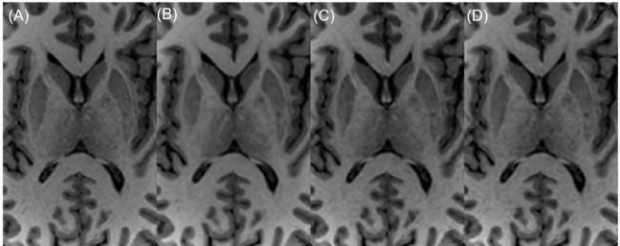

Figure 4. Example image slices, showing the deep gray matter, reconstructed by (A) ARC (net acceleration factor R=2.1), (B) DL-Speed with 2D Conv (net R=10), (C) DL-Speed with alternating 2D Conv (net R=10), and (D) DL-Speed with 3D Conv (net R=10), evaluated on retrospectively-undersampled 3D MPRAGE brain data.

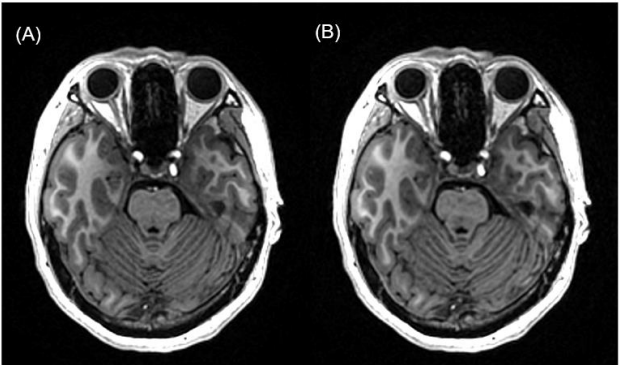

Figure 5. Example image slices reconstructed by (A) ARC with net acceleration factor R=2.1 (scan time: 5.1 min) and (B) DL-Speed with 2D Conv with net R=10 (scan time: 1.1 min), evaluated on prospectively-undersampled 3D MPRAGE brain data. The slices in (A) and (B) do not necessarily match because they are from different prospective scans.

|  | DL-Speed with 2D Conv | DL-Speed with alternating 2D Conv | DL-Speed with 3D Conv |
|---|---|---|---|
| nMSE (%) | 0.20 ± 0.025 | 0.26 ± 0.030 | 0.24 ± 0.048 |
| SSIM | 0.987 ± 0.002 | 0.989 ± 0.001 | 0.988 ± 0.001 |

Table 1. Magnetization as a function of applied field. Note that "Fig." is abbreviated. There is a period after the figure number, followed by two spaces. It is good practice to explain the significance of the figure in the caption. Comparison of normalized mean-squared error (nMSE) and SSIM for DL-Speed variants with 2D Conv, alternating 2D Conv and 3D Conv evaluated on 5 retrospectively-undersampled 3D T1-weighted MPRAGE brain datasets with a 10-fold acceleration. The table shows mean metrics across the 5 data sets with standard deviation.

|  | Zero-filled | Compressed sensing | DL-Speed |
|---|---|---|---|
| nMSE (%) | 8.19 ± 1.11 | 3.52 ± 0.73 | 1.01 ± 0.61 |
| SSIM | 0.755 ± 0.024 | 0.833 ± 0.029 | 0.926 ± 0.034 |

Table 2. Comparison of normalized mean-squared error (nMSE) and SSIM for zero-filled, total-variation-based compressed sensing, and DL-Speed with 2D Conv evaluated on 15 retrospectively-undersampled 3D T1-weighted abdominal scan datasets with a mean acceleration factor of 9.4 (standard deviation 0.3). The table shows mean metrics across the 15 data sets with standard deviation.

Figure 5 shows example image slices reconstructed from prospectively-undersampled MPRAGE data, by the standard ARC with R=2.1 (scan time: 5.1 min) and DL-Speed with R=10 (scan time: 1.1 min). They were assessed as having equivalent image quality in all 3 volunteers.

### B. Evaluation on LAVA Scan Data

Table 2 compares DL-Speed to zero-filled and the total variation based compressed sensing[33], evaluated on 15 retrospectively-undersampled 3D T1-weighted LAVA abdominal scan datasets with mean R=9.4, regarding normalized MSE and SSIM. DL-Speed outperformed the compressed sensing in the quantitative metrics. Figure 6 shows example images for the ground truth and for zero-filled, total variation based compressed sensing and DL-Speed images reconstructed from retrospectively-undersampled 3D LAVA data (R=9.2).

Figure 7 shows example image slices reconstructed from prospectively-undersampled breath-hold 3D LAVA scan data by the standard ARC (Figure 7A) with a net acceleration factor of R=3.8 (scan time: 11.8 s) and DL-Speed (Figure 7B) with R=9.3 (scan time: 4.9 s). Reconstructed images from free-breathing scans with a navigator-based respiratory gating are also shown for ARC (Figure 7C) and DL-Speed (Figure 7D).

### C. Reconstruction Time

DL-Speed inference implemented using TensorFlow was tested on a GPU (GeForce RTX 3090 with 24 GB memory, Nvidia). The reconstruction time for generating a brain image volume (e.g., Figure 5B) from undersampled MPRAGE data (R=10) with 256 frequency encodings, 232 phase encodings, 190 slices and 32 channels took about 93 s (72 s for inferencing) for DL-Speed with 2D Conv (28 iterations, 9 layers per iteration, 96 filters per layer, and 20 skip connections).

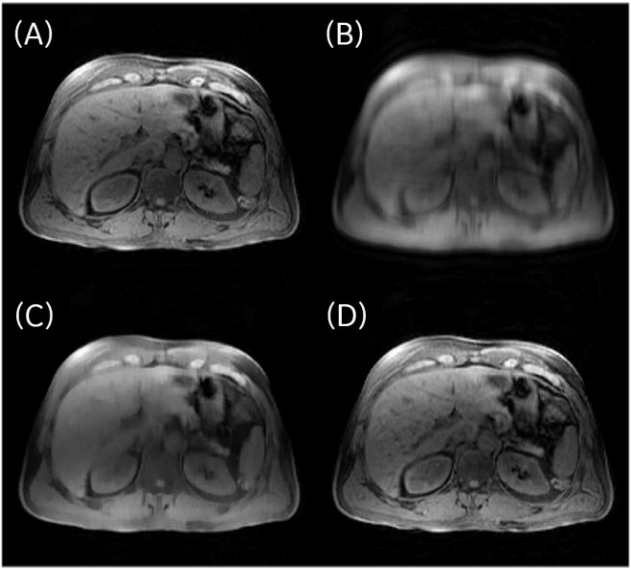

Figure 6. Example image slices reconstructed from retrospectively undersampled 3D LAVA data with a 9.2-fold acceleration: (A) ground-truth fully-sampled, (B) zero-filled, (C) total-variation based compressed-sensing, and (D) DL-Speed with 2D Conv.

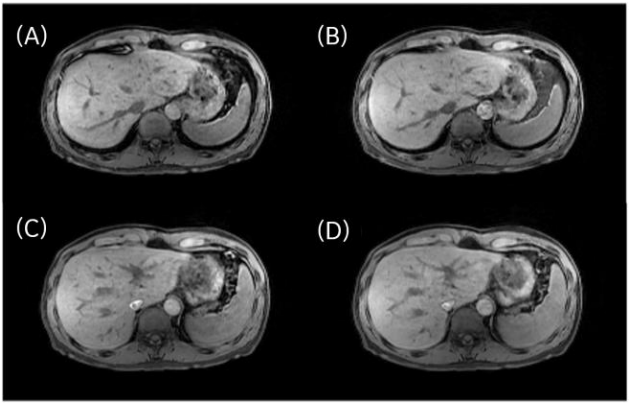

Figure 7. Example image slices reconstructed from prospectively undersampled 3D LAVA abdominal data: breath-hold scan for (A) ARC with a net acceleration R=3.8 and regular sampling (scan time: 11.8 s), and (B) DL-Speed with 2D Conv with R=9.3 and VDPD sampling (scan time: 4.9 s); and free-breathing scan with respiratory gating based on a pencil-beam navigator tracker for (C) ARC with a net acceleration R=3.8 and regular sampling (estimated scan time: 41.7 s), and (D) DL-Speed with 2D Conv with R=9.3 and VDPD sampling (estimated scan time: 16.4 s). The slices in (A) and (B) do not necessarily match because they are from different prospective scans. Neither do the slices in (C) and (D).

## IV. Discussion

In this study, we evaluated DL-Speed, an unrolled optimization architecture, on various datasets. The scores received from radiologists by DL-Speed with a 10-fold acceleration were higher than or equal to those by the standard parallel imaging with a 2.1-fold acceleration for categories of SNR, artifacts, gray/white matter contrast, resolution/sharpness, deep gray, cerebellar vermis, anterior commissure, and overall quality, based on retrospectively-undersampled 3D MPRAGE data (Figure 2). This implies DL-Speed can achieve up to a 5-fold acceleration compared to conventional parallel imaging for 3D MPRAGE scan without compromising image quality significantly while enabling real-time applications (e.g., 1.6 min for reconstructing a brain image volume on a GPU). For example, it was demonstrated that a scan time of 5.1 min is reduced to 1.1 min in a prospective MPRAGE scan study (Figure 5). In the retrospective study, DL-Speed received higher scores than the standard parallel imaging, which was considered as the ground truth (Figure 2). This is consistent with a previous 2D knee study[17] where readers preferred DL-accelerated images than standard clinical images. A net 10-fold acceleration for 3D MRI we demonstrated is also consistent with previous 3D studies.[6, 10, 18] We have shown a proof-of-concept but more extensive comparison will need to be performed to assess the trade-off between acceleration and image quality. We note that DL-Speed can perform denoising as a regularizer in addition to sparse reconstruction and it will be worthwhile to investigate the denoising aspect of DL-Speed.

The DL-Speed trained on brain data generalized reasonably well to 3D LAVA abdominal scan data (Figures 6-7, and Table 2). It was demonstrated that DL-Speed can accelerate 3D LAVA abdominal scans by a factor of 9-10, leading to about a 2- to 3-fold acceleration compared to conventional parallel imaging. In fact, it is not easy to build good training datasets based on abdominal data, which tend to be of lower quality in terms of motion artifacts, B0/B1 inhomogeneity, resolution, and overall SNR. The generalizability of unrolled optimization methods may come from the fact that the methods rely not only on neural network-based regularization but also on data consistency. For example, the DL-Speed trained on brain data was also shown to generalize to arterial spin labeling based renal perfusion.[34] However, it will still be worthwhile to train a network using good abdominal training datasets, or a small number of good abdominal datasets with fine-tune training, and evaluate it to see if there is any improvement in performance.

In 3D imaging, it is natural to use 3D filters in convolutional layers to exploit correlations between adjacent slices. However, DL-Speed using 2D filters outperformed DL-Speed using 3D filters (Table 1 and Figure 2). This is probably because the DL-Speed with 3D Conv as well as DL-Speed with alternating 2D used fewer iterations, filters, and skip connections due to the memory limitation. Given that 3D filters have a potential for better performance than 2D filters, we may revisit DL-Speed with 3D Conv when hardware/software advances permit a deeper architecture.

To train DL-Speed, we used a contrast-weighted SSIM loss function, which yielded sharper and less blurred images than L1 loss function.[26] In fact, we extended the contrast-weighted SSIM to complex-valued images (note conventional SSIM usually only applies to real-valued nonnegative images such as magnitude images). It makes sense to make a loss function phase-aware given that MR images are intrinsically complex-valued. However, we did not observe a noticeable difference in results between the contrast-weighted SSIM loss functions for magnitude images and complex-valued images. This is probably because the data consistency already deals with most of phase-pertinent information in the applications we considered. It will be worthwhile to investigate the phase-aware

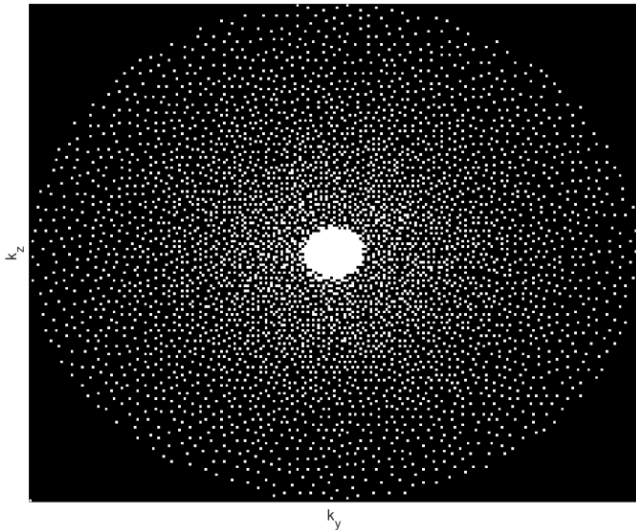

Supporting Figure S1. Example variable-density Poisson-disc (VDPD) sampling pattern with a 10-fold acceleration, a fully-sampled central region, and corner cutting in a 2D phase-encode ($k_y$) and slice-encode ($k_z$) plane for 3D MRI.

loss function in phase-based applications such as fat-water separation and flow quantification.[19]

## V. CONCLUSION

DL-Speed was demonstrated to accelerate 3D MRI scans such as MPRAGE and LAVA with up to a net 10-fold acceleration, achieving a 2- to 5-fold acceleration compared to conventional parallel imaging while maintaining diagnostic image quality and enabling real-time reconstruction. The DL-Speed reconstructed images were at least equivalent to conventionally-accelerated brain images but with a scan that is 5-times shorter. DL-Speed trained using brain scan data generalized to LAVA abdominal scan data. DL-Speed has a potential to accelerate other sequences and applications such as multiphase dynamic contrast enhanced MRI.

## ACKNOWLEDGEMENT

The authors would like to thank Xucheng Zhu, Marc Lebel, Ersin Bayram, Suchandrima Banerjee and Anja Brau for helpful discussions and comments on this study.